# The potential of Rutherford Backscattering Spectrometry for composition analysis of colloidal nanocrystals

D. Primetzhofer[(1)], M. Sytnyk[(2)], P. J. Wagner[(3)], P. Bauer[(3)], W. Heiss[(2)]

[(1)] Ion physics, Department of Physics and Astronomy, Uppsala University, 75120 Uppsala, Sweden

[(2)] Institute of Semiconductor and Solid State physics, University Linz, Altenbergerstrasse 69, 4040 Linz, Austria

[(3)] Institute of Experimental physics, University Linz, Altenbergerstrasse 69, 4040 Linz, Austria

**Abstract:**

We investigate the potential of elastic scattering of energetic ions for compositional analysis of magnetic colloidal nanocrystals. Thin layers of nanocrystals deposited by spin coating on Si-wafers are investigated by two different set-ups for Rutherford Backscattering Spectrometry (RBS), employing different projectile ions ($^4$He, $^{12}$C) and primary energies (600 keV – 8 MeV). The advantages and disadvantages of the different approaches are discussed in terms of obtainable mass resolution, necessary primary particle fluence and deposited energy. It is shown that different isotopes of transition metals can be resolved by employing 8 MeV $^{12}C^{3+}$ primary ions.

## 1. Introduction:

Nanocrystals with complex composition and structure have emerged from a material class of fundamental interest to industrial relevance [1]. Postsynthetic modification of composition and structure of nanocrystals permits to tailor their electrical or magnetic properties in ways not practicable by direct synthesis (see e.g. [2],[3]). Along such research lines a reliable composition analysis of the nanocrystalline material is demanded. An ideal analysis method for composition measurements should be able to operate with small amounts of sample material and permit unique identification of the constituents and determination of their respective concentrations. The analysis should be furthermore independent of the size and the size distribution as well as the organic coating or solvent in case of colloidal nanoparticles.

Elastic scattering techniques employing energetic ions with energies of several hundred keV per atomic mass unit can fulfill these requirements, since they yield information on the composition via scattering from the atomic nucleus, free from matrix effects, and for sample amounts of less than a microgram (see e.g. [4] for an overview).

Due to these properties, several recent studies explored and used the potential of elastic scattering - mainly the Medium-Energy Ion Scattering (MEIS) methodology - for analysis of colloidal nanocrystals [5],[6],[7]. As many of these studies demonstrate, enhanced scattering cross sections, and thus also lower sample material amounts necessary for the analysis can be achieved e.g. by employing low primary energies of around 100 keV. This, however, demands drastically increased energy resolution to keep up analytical performance. This can be achieved e.g. by use of magnetic or electrostatic spectrometers [8],[9].

We present an investigation of the composition of colloidal nanocrystals with different compositions by Rutherford Backscattering Spectrometry (RBS) employing different ion species at a number of different primary energies. In one approach, the primary ion energy is reduced for an increased scattering cross section and a high-resolution surface barrier detector is employed [10]. In the other approach heavier ions with at energies of several MeV are employed as projectiles to achieve superior mass resolution, necessary for characterization of a system of magnetic nanocrystal heterostructures [11]. We give a detailed description of the experimental details, the evaluation and the achievable accuracy as well as how to avoid complications that may arise in such a kind of analysis.

## 2. Experiment:

The ion scattering experiments in Linz were performed using the AN-700 van de Graaff accelerator at the Department of Atomic Physics and Surface Science of the JKU Linz. The accelerator is capable of creating monoenergetic beams of H, D atomic and molecular ions and $He^+$ ions with energies in the range 30 – 700 keV and primary beam currents up to several µA. In the present study 600 keV $He^+$ ions were used due to their favourable mass resolution. The samples were analyzed in a HV-chamber with a base-pressure of $1\cdot10^{-5}$ Pa.

Scattered ions were detected by two semiconductor surface barrier detectors. The first detector, situated in 150.1° in Cornell geometry features very high energy resolution (FWHM < 1.8 keV). This is obtained by $LN_2$ cooling of the detector and of the input stage of the preamplifier, and by minimizing the input capacitance of the set-up. Furthermore, the housing of the amplifying FET is optimized to keep it at optimum temperature (~ 100 K) and to avoid microphonic noise from mechanical vibrations [10]. The electronic noise of this detection system is found ~ 500 eV from detection of soft x-rays. The main contribution that decreases energy resolution for ions is energy loss straggling in the dead layer of the detector which finally yields an energy resolution of 1.8 keV for protons and about 7 keV for He-ions. By use of this detector high-resolution depth profiling experiments of nm-films [12] and analysis of samples with high demands on mass resolution as identification of trace elements in organic samples [13] has been performed successfully. The second detector, situated in 154.6° in IBM geometry features a larger solid angle and thus higher count rate, but lower energy resolution.

The experiments in Uppsala were performed employing a 5 MV 15SDH-2 pelletron tandem-accelerator at the Ångström laboratory. By the use of different ion-sources beams of many different ion species with energies in the range of 2 – 50 MeV can be produced. In the present study 2 MeV $^4He^+$ ions and 8 MeV $^{12}C^{3+}$ ions were used. The scattering chamber with a base-pressure of < $1\cdot10^{-5}$ Pa holds two solid state detectors for RBS, in Cornell geometry, with the first situated fixed at 170.0° while the second can be positioned at almost arbitrary scattering angles.

Two different systems of nanocrystals were investigated. The first set (system I) were CdSe- and CdS-nanocrystals for which by cation-exchange Cd was exchanged via Hg to Mn. System II was formed from iron-oxide nanocrystals for which cation-exchange was employed to directly exchange Fe by Co [11].

The nanocrystalline samples were kept in high-purity chloroform solution after synthesis. In order to permit a unique identification of present elements by scattering kinematics thin films [14] of nanocrystals were prepared from the solutions. To obtain compositional information, a substrate with atomic number lower than the constituents is desirable. Deposition on a single crystalline Si-wafer was chosen, since it features a high-quality surface and permits identification of all constituting elements in system I and the metal components of system II, respectively. Deposition was done via spin coating at different rotation speeds. The aim was on achieving a thin homogenous layer of nanocrystals in order to perform quantitative analysis. Fig. 1 shows simulations for a typical RBS spectrum obtained for a high-quality thin film (red line), in comparison to a film featuring a broad thickness distribution (black line). For quantification of concentrations simulation of RBS-spectra was performed using the SIMNRA simulation package [15].

As can be readily deduced from the figure, a broad distribution of nanocrystals clearly hampers the analysis of the composition, especially for elements adjacent in the periodic table. Significant clustering can be expected in case of deposition of multilayers of nanocrystals. To compare, a monolayer of nanocrystals with typical diameters of about 5 nm corresponds to an areal density of ~25 $10^{15}$ atoms/cm$^2$. Since significant clustering is expected for the equivalent of a monolayer, the aim is to achieve sub-monolayer coverage, which in turn, however, means a low but narrow signal, as can be seen from comparison to the substrate signal in Fig. 1.

### 3. Results and discussion:

The analysis of samples for system I was performed mainly by the set-up in Linz and for several samples also in Uppsala. Experiments in Linz were performed using a 600 keV $^4$He$^+$ primary ion beam. Reference experiments on the Si-wafers showed the absence of contaminations in the substrate. A typical spectrum resulting from an experiment using a spin coated thin film of crystals from system I is shown in Fig. 2. As can be seen, distinct peaks can be identified, each corresponding to scattering from a certain element in the sample. From fitting the primary ion current and the solid angle of the detector by the Si background spectrum the apparent thickness of the sample is found to be 10·$10^{15}$ atoms/cm$^2$, which corresponds to a coverage well below one monolayer. Some minor channelling effects in the Si can be observed in the shape of the spectrum, which indicates that the coverage might be even lower in consequence of the reduced Si yield. The Si-signal is also slightly affected by pile-up in the detection system which forms a tail towards higher energies.

The accuracy of the evaluation is only affected in a significant way by two contributions: the major contribution is due the experiments statistics. When giving the relative concentrations of the constituents, as in this case Mn, Se and Cd, the error is dominated by the element featuring lowest peak area, i.e. Mn with about 250 counts. The resulting relative error of about 6% in Mn concentration of course also affects the concentrations of the other constituents, weighted with their contribution to the spectrum, resulting in the present case in about a twice as strong effect of the Mn error on the uncertainty in the Se-concentration than on the Cd-concentration. However, the relative ratio of Cd and Se can, from the same spectrum, be determined with much higher precision due to higher scattering cross sections and concentrations and is found to be 1.02 +/- 0.025. This ratio was also subsequently checked for the same sample using the set-up in Uppsala and 2 MeV $^4$He$^+$ ions with a ratio of 1.05 +/- 0.1, with the larger error due to the worse statistics of the recorded spectrum (not shown). A smaller possible systematic error is due to the scattering cross section. Cross sections for all presented experiments cannot be described exclusively by the Rutherford cross section due to screening by bound electrons. This effect is found well below 3% for any of the present experiments and is accounted for in the simulations by the correction suggested by Andersen et al. [16]. Thus, the maximum systematic error is expected to be much lower than 1% and thus the statistical one dominates in all the experiments.

For the analysis of system II, increased requirements are found regarding the energy resolution necessary to resolve iron and cobalt. For scattering of protons at classical MEIS conditions with a scattering angle of 120 deg and 100 keV protons as primary beam, the energy separation is as low as 250 eV. Even when helium would be used as projectile, the separation stays below 1 keV. When the energy loss due to the spatial extension of the nanocrystals (~25 $10^{15}$ atoms/cm$^2$) is considered, values of about 500 eV and 750 eV with a corresponding straggling distribution for H and He ions respectively, have to be expected. Since the organic coating of the nanocrystals will add another contribution to the energy loss it becomes obvious, that low ion energies may not form a straightforward option. Instead, increasing the ion energy can also be used to increase the mass separation. Both, 4 MeV $^4$He and 8 MeV $^{12}$C yield energy separations in a 170° backscattering geometry which are found with 35 keV and 120 keV about twice the expected energy resolution of a solid state detector at the respective detection energies [17]. In terms of the (Rutherford-) scattering cross section $^{12}$C at 8 MeV is found twice as high than 4 MeV $^4$He. Also C does not undergo any nuclear reactions at these energies with the substrate Si, as seen for

$^4$He [18]. At the same time, the deposited energy due to nuclear and electronic stopping is found significantly higher for C. Fig. 3 presents experimental spectra as well as simulations by SIMNRA for scattering of 8 MeV $^{12}$C$^+$ from sub-monolayer films of nanocrystals a) before and b), c) after the cation-exchange reaction. The spectrum shown in a) indicates, that the quantum dots before the exchange-reaction contain no detectable Co. The absence of an exponential tail towards lower energies shows the absence of clustered nanocrystals on the surface. In fact, the calculated areal density is again found to be ~10·10$^{15}$ atoms/cm$^2$, in accordance with sub-monolayer coverage. Instead, the slight asymmetry in the peak associated with scattering from Fe can be attributed to the 6% natural abundance of the isotope $^{54}$Fe relative to the larger fraction of $^{56}$Fe. This is also more clearly detectable in the other spectra b) and c). In Fig. 3 b), after the cation-exchange, Fe and Co signals can be clearly separated. Also a contamination by Cl from the CoCl$_2$ in the cation exchange reaction (for details see [11]) can be readily detected. Fig. 3 c) shows a nanocrystal sample for which the exchange reaction was performed for which the residual chlorine from the reaction was removed successfully by a repeated washing procedure.

Note that, for all presented investigations, time resolved spectra did not show any observable influence of irradiation time on the deduced concentrations, for the employed ion dose. This is in accordance with expectations that in the case of an unlikely but possible depletion of Fe and Co in the sample due to the energy deposition both species would be affected in a similar way due to their similar mass and chemistry.

## 4. Summary and Conclusions:

In this contribution we presented how elastic scattering, i.e. classical RBS can be employed to characterize small amounts of colloidal nanocrystals in order to quantify compositional changes during cation-exchange reactions. We showed that a straightforward quantitative evaluation of the experiment is possible, if samples are reasonably prepared. The fact that the experiments are performed at energies in the single scattering regime exclusively and that stopping powers are not relevant is the foundation for the analytical simplicity of the approach. Besides the experimental statistics, only very minor (<<1%) uncertainties due to scattering cross sections in the simulations enter the results.

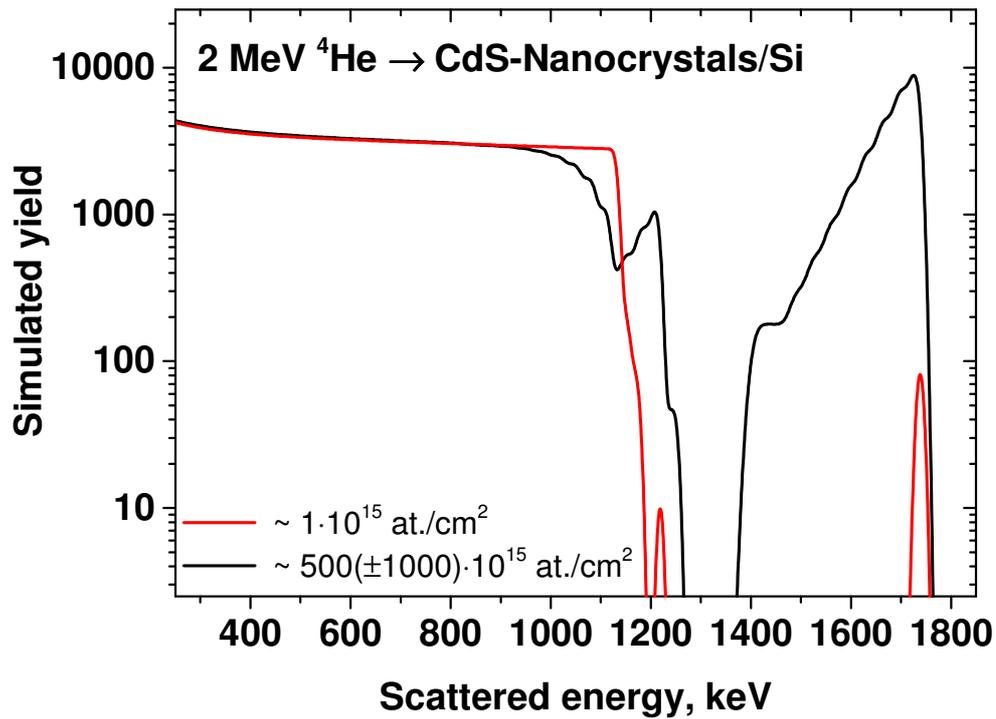

Fig.1: Simulated RBS spectra by SIMNRA [15] for 2 MeV $^4$He$^+$ ions scattered from thin films of CdS on a Si substrate. For a sub-monolayer film, without clustering of the crystals a sharp peak structure is expected (red line). The nanocrystals will tend to form clusters on the surface, which will lead to spectra as shown by the black curve. The exponentially decaying low energy tails of the spectra can complicate analysis, especially, when the interest is in species of similar mass.

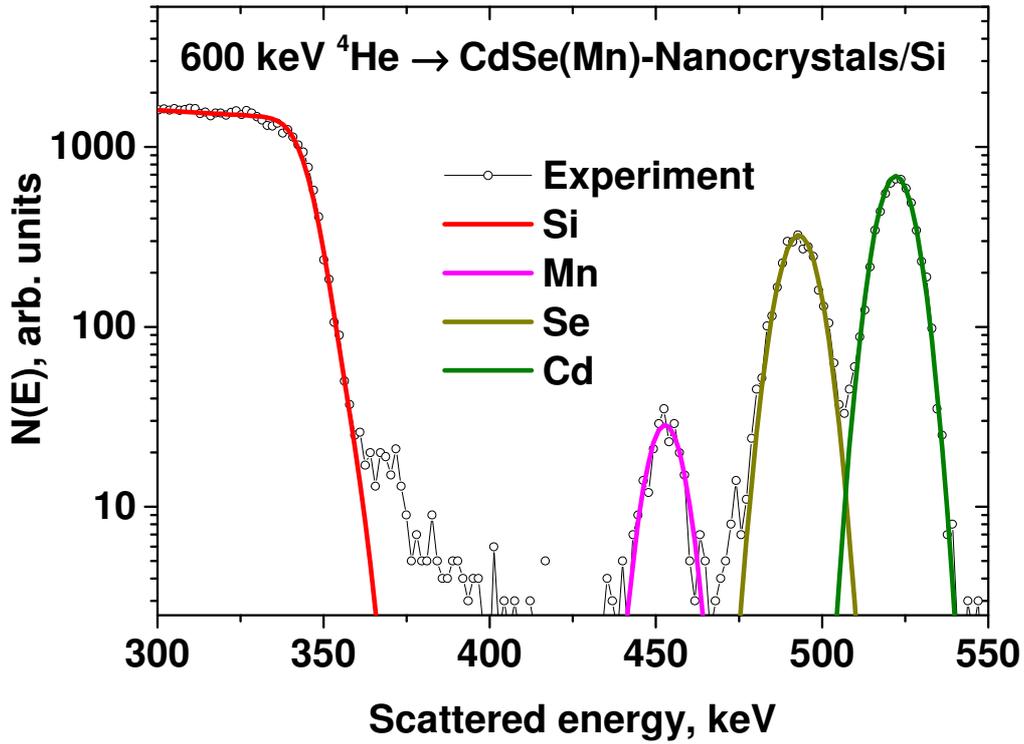

Fig.2: Experimental RBS spectra for 600 keV $^4$He$^+$ ions scattered from thin films of CdSe nanocrystals in which part of the Cd was exchanged via Hg to Mn (open symbols). The films are spin-coated on a Si substrate. The film exhibits good quality, the Si-signal is slightly affected by pile-up in the detection system. Also shown is a simulation by SIMNRA [15] to quantify the concentrations.

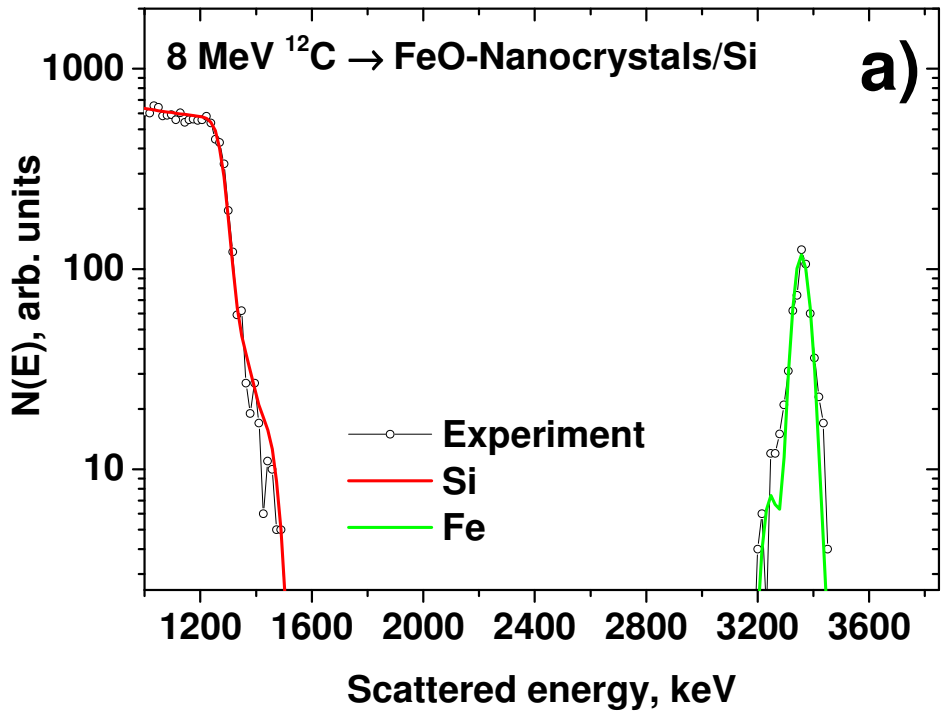

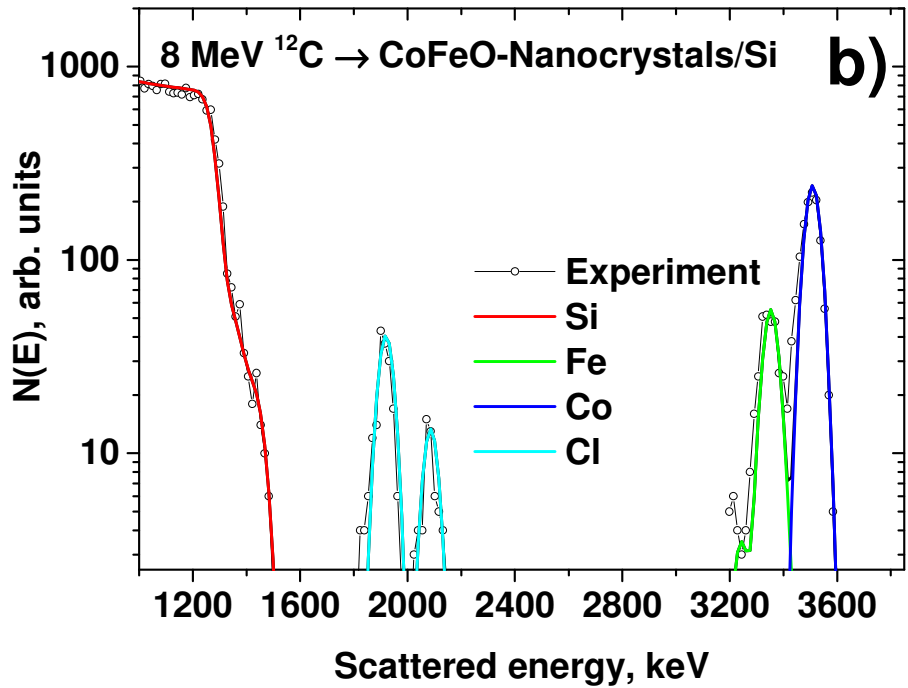

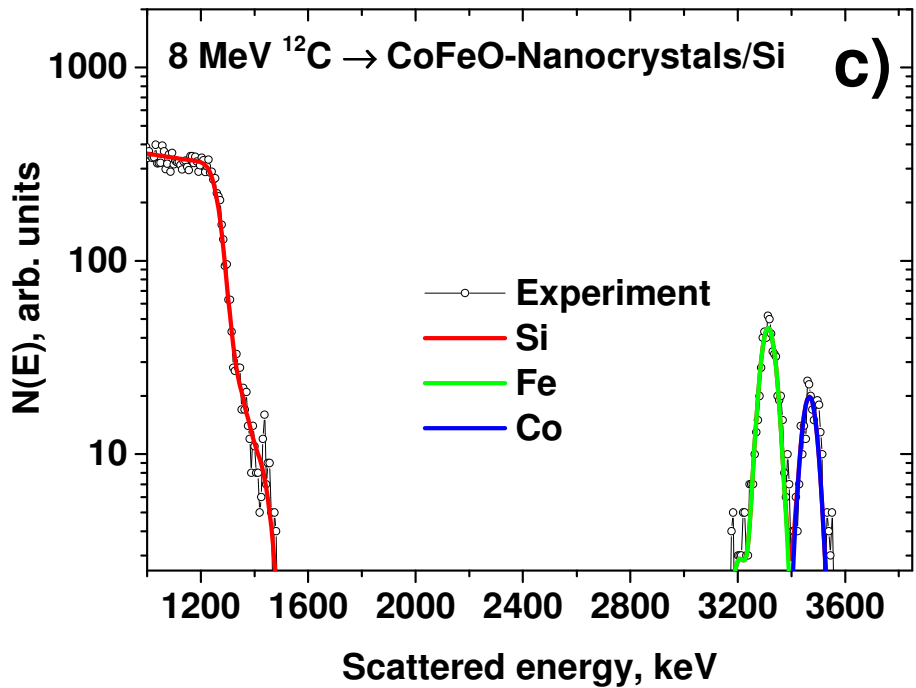

*Fig.3: Experimental RBS spectra (open symbols) and simulations by SIMNRA [15] (full lines) for sub-monolayer films of colloidal nanocrystals on Si. a) before cation-exchange b), c) after cation-exchange with and without residual $CoCl_2$ from the reaction. The signals for the adjacent elements Fe (green) and Co (blue) can be clearly separated. Also isotopes of chlorine (light blue) from the exchange reaction present in accordance with their natural abundance are readily distinguished.*